\documentclass{pasj00}
\newcommand\lsim{\lower0.5ex\hbox{$\; \buildrel < \over \sim \;$}}
\newcommand\gsim{\lower0.5ex\hbox{$\; \buildrel > \over \sim \;$}}

\draft

\begin{document}
\SetRunningHead{Chen, Li, \& Wang}{LMXB as the Progenitor of PSR
J1713+0747} \Received{2005/09/27}{} \Accepted{2005/12/22}{}

\title{ Low-Mass X-Ray Binary as the Progenitor of PSR J1713+0747}

\author{W. C. \textsc{Chen}, X. D. \textsc{Li}, and Z. R. \textsc{Wang}}
\affil{Department of Astronomy, Nanjing University, Nanjing
210093, China} \email{chenwc, lixd, zrwang@nju.edu.cn}


%


\maketitle

\begin{abstract}
We ¢ô have calculated the evolution of low-mass X-ray binaries
that lead to the formation of the binary radio pulsars, like PSR
J1713+0747. We showed that the mass transfer is most likely to be
nonconservative, due to unstable disk accretion, to account for
the mass of PSR J1713+0747, which is close to its initial value.
We assumed that part of the lost material from the binary may form
a circumbinary disk, and found that it can significantly influence
the mass-transfer processes. We briefly discuss the implications
of the circumbinary disks on the evolution of low-mass X-ray
binaries and the formation of low-mass binary pulsars.
\end{abstract}

\section{Introduction}

Since the first radio pulsar was discovered by \citet{hewi68}, the
number of pulsars has already increased to be more than 1500. In
this population only $\sim$ 3\% are the members of binary systems.
Most of the binary radio pulsars are millisecond pulsars
(\cite{back91}) with a He or CO white-dwarf companion. Such binary
and millisecond pulsars are thought to have been ``recycled" from
accreting low-mass or intermediate-mass X-ray binaries (L/IMXBs,
see \cite{bhat91,tau06} for reviews). Mass accretion onto a
neutron star induces magnetic field decay, and spins the star up
to a short period. When mass transfer ceases, the end-point of the
evolution is a circular binary containing a neutron star visible
as a low-field, millisecond radio pulsar, and a white dwarf, the
remaining He or CO core of the companion. The recent discovery of
millisecond accreting pulsars has lent strong support to this
scenario \citep{vk06}.

The mechanisms that drive mass transfer in LMXBs depend on the
initial separations of the binary components (\cite{bhat91}). In
narrow systems with initial orbital periods of $P_{\rm orb}< 1-2$
d, mass transfer is driven by the loss of orbital angular momentum
via gravitational radiation and/or magnetic braking. Mass transfer
in relatively wide ($P_{\rm orb}\gsim 1-2$ d) LMXBs is driven by
the nuclear expansion of the secondary. Systems of this kind form
a quite homogeneous group whose evolutionary history seems to be
well understood (\cite{webb83}; \cite{taam83}). \citet{ritt99}
derived a simple analytical solution for the evolution of a close
binary with nuclear time-scale driven mass transfer from a giant,
based on the well-known fact that the luminosity and the radius of
a giant scale to a good approximation as simple power laws of the
mass of the degenerate helium core. It has been shown that the
orbital binary periods gradually increase during evolution. Once
the initial orbital period, $P_{\rm orb}$, is long enough, the
mass transfer rate, $\dot{M}$,  will exceed the Eddington
accretion rate, $\dot{M}_{\rm E}\simeq 1.5\times
10^{-8}\,M_{\odot}{\rm yr}^{-1}$, of a neutron star, probably
leading to mass loss, and the mass transfer becomes
nonconservative.

The 4.57 ms radio pulsar PSR J1713+0747 is in a 67.8 day circular
orbit with a low-mass white-dwarf companion (\cite{fost93}). These
characteristics indicate that this pulsar probably had evolved
from a LMXB consisting of a neutron star and a low-mass red giant.
Recent observations \citep{spla05} constrain the masses of the
pulsar and secondary star to be $(1.3\pm 0.2)\,M_{\odot}$ and
$(0.28\pm 0.03)\,M_{\odot}$, respectively (68\% confidence),
implying extensive mass loss during the previous LMXB evolution.

In this paper we explore nonconservative mass transfer in the
LMXBs evolution, leading to systems like PSR J1713+0747, while
taking into account mass loss resulting from unstable mass
transfer, as well as its feedback to the binary evolution. In
section 2 we describe our model for the evolution of LMXBs. In
section 3 we present the calculated results, and compare them with
the observational data of PSR J1713+0747 in section 4. We present
a brief discussion and conclude in section 5.

\section{Model}
In a binary system consisting of a neutron star (of mass $M_{\rm
X}$) and a low-mass red giant (of mass $M$), the evolutionary
expansion of the Roche lobe-filling giant will transfer mass to
the neutron star through the inner Lagrangian point on the nuclear
time scale. The mass transfer rate can be described by the
following equation (e.g., \cite{ritt99}):
\begin{equation}
\dot{M}=-\left( \frac {\dot{R}}{R} \right) _{M}\frac{M}{\xi_{\rm
e}-\xi_{\rm R}},
\end{equation}
where $\xi_{\rm e}=(\partial \ln R/ \partial\ln M)_{\rm eq}$ is
the thermal equilibrium mass radius exponent and $\xi_{\rm
R}=\partial\ln R_{\rm L}/ \partial\ln M$ is the mass radius
exponent of the Roche lobe radius of the giant (\cite{sobe97});
$R$ and $R_{\rm L}$ are the radii of the giant and of the Roche
lobe, respectively.

\subsection{Nuclear Evolution of the Giant}
 For a lower giant-branch star, the stellar luminosity, $L$, and
radius, $R$, are uniquely determined by the mass of its degenerate
helium core, $M_{\rm c}$, independent of the hydrogen envelope
mass (\cite{refs70}),
\begin{equation}
 L=L_{\odot}\exp(\Sigma a_{i}y^{i}),
 \end{equation}
\begin{equation}
R=R_{\odot}\exp(\Sigma c_{i}y^{i}),
\end{equation}
where $y=\ln(4M_{\rm c}/M _{\odot})$ and $i=0$, 1, 2, 3. The
parameters $a_{i}$ and $c_{i}$ in the above equations for the
solar chemical composition can be found in Webbink (1975).

From equation (3), the change in the radius of the donor star is
related to the growth rate, $\dot{M_{\rm c}}$, of the helium core
by
\begin{equation}
 \left(\frac {\dot{R}}{R} \right) _{M}
 =(c_{1}+2c_{2}y+3c_{3}y^{2})\frac{\dot{M_{\rm c}}}{M_{\rm c}}.
 \end{equation}

Because the stellar luminosity is directly related to the growth
in the core
 mass due to shell hydrogen burning,
\begin{equation}
 L=X\varepsilon_{\rm H}\dot{M_{\rm c}},
 \end{equation}
where $X$ is the initial hydrogen mass fraction in the shell, and
$\varepsilon_{\rm H}\simeq6\times10^{18}\,{\rm erg\quad g}^{-1}$
is the released energy rate by the CNO cycle (\cite{webb83}).

Since the giant remains in thermal equilibrium during the mass
transfer, and its radius is only determined by the core mass,
independent of total mass, $M$, $\xi_{\rm e} \approx 0$
(\cite{sobe97}). We then have
 \begin{equation}
 \dot{M}=\frac{c_{1}+2c_{2}y+3c_{3}y^{2}}{\xi_{\rm R}}(\frac{M}{M_{\rm c}})\dot{M_{\rm c}}.
 \end{equation}

\subsection{Mass and Angular Momentum Loss}
The orbital evolution is determined by the angular-momentum loss
rate for a binary system possessing mass communion. The total
angular momentum of a system with a circular orbit is
\begin{equation}
J=a^{2}\frac{MM_{\rm X}}{M+M_{\rm X}}\frac{2\pi}{P_{\rm orb}},
\end{equation}
where $a=[G(M+M_{X})P_{\rm orb}^{2}/(4\pi^{2})]^{1/3}$ is the
binary separation; $G$ and $P_{\rm orb}$ are the gravitational
constant and the orbital period of system, respectively. We
neglect the spin angular momentum of the components because it is
small compared the total orbital angular momentum of the system,
and consider the following processes for the mass and orbital
angular momentum loss from the binary system.

\noindent{\em 1. isotropic wind}

We consider the mass and angular momentum loss during
nonconservative mass transfer in the evolution of LMXBs. If the
magnitude of the secular mass transfer rate, $-\dot{M}$, is
greater than the Eddington accretion rate, $\dot{M}_{\rm E}$, we
assume that the neutron star accretes a fraction $(1-\alpha)$ of
the transferred mass, and the remaining fraction, $\alpha$, is
ejected in the vicinity of the neutron star ($\alpha=0$ if
$|\dot{M}|\leq\dot{M_{\rm E}}$,
 else $\alpha=1+\dot{M}_{\rm E}/\dot{M}$).

Mass loss can also occur when the mass transfer is sufficiently
low. It is well known that disk accretion in LMXBs is thermally
and viscously unstable if the mass-transfer rate is less than a
critical value, $\dot{M}_{\rm cr}$ (van Paradijs 1996;
\cite{dubu99})
\begin{equation}
\dot{M}_{\rm cr} \simeq 3.2\times10^{-9}M_{\odot}{\rm
yr}^{-1}\left(\frac{M_{\rm
X}}{1.4M_{\odot}}\right)^{0.5}\left(\frac{M}{1M_{\odot}}\right)^{-0.2}\left(\frac{P_{\rm
orb}}{1 \rm d}\right)^{1.4}.
\end{equation}
When an accretion-disk instability occurs, the accreting neutron
star will become a transient X-ray source, experiencing outbursts
separated by long quiescent intervals. For the accretion behavior
during outbursts, we adopt the following prescription suggested by
Portegies Zwart et al. (2004). The accretion rate, $\dot{M}_{\rm
X}$, first reaches a peak value of $2\dot{M}_{\rm E}$, and then
decays with an exponential timescale of $t_{\rm d}=6\,{\rm
d}\times \min(1,\,P_{\rm orb}/10\,{\rm hr})$. The relation of the
recurrence time, $t_{\rm r}$, and the total energy, $E$, in an
outburst satisfies $t_{\rm r}=E/(-0.1\dot{M}c^{2})$, where
$\log(E/{\rm erg})=45+\log (P_{\rm orb}/{\rm d})$. Since the
accretion rate of a neutron star is $\dot{M}_{\rm X}=2\dot{M}_{\rm
E}\exp(-t/t_{\rm d})$, we can calculate the accreted mass of the
neutron star within the recurrence time, $t_{\rm r}$, to be
\begin{equation}
\Delta M_{\rm X}=2\dot{M}_{\rm E}t_{\rm d}[1-\exp(-t_{\rm
r}/t_{\rm d})].
\end{equation}
Accordingly, the fraction $\alpha=1+\Delta M_{\rm
X}/(\dot{M}t_{\rm r})$ of the transferred mass is assumed to be
lost from the system.

The material probably leaves the binary system in the form of
winds, outflows, or jets. We assume that a fraction of
($1-\delta$) of the lost matter is ejected in the vicinity of the
neutron star, carrying away a specific orbital angular momentum of
the neutron star,
\begin{equation}
j_{\rm w}=\frac{M}{M_{\rm T}M_{\rm X}}J,
\end{equation}
where $M_{\rm T}$ is the total mass of the system.

\noindent{\em 2. CB disk}

Spruit and Taam (2001) suggested that a Keplerian, circumbinary
(CB) disk could be formed as the result of mass outflow from an
accreting star or accretion disk. Tidal torques are put on the CB
disk,
 and carry away angular momentum from the binary orbiting
 inside it. We assume that the other part (with a
fraction of $\delta$) of mass loss forms a CB disk, extracting the
orbital angular momentum from the binary system.

At the inner edge, $r_{\rm i}$, of the disk, the viscous torque
exerted by the CB disk on the binary can be shown to be
(\cite{spru01,taam01})
\begin{equation}
T=\gamma(\frac{2\pi a^2 }{P_{\rm
orb}})\delta\dot{M}(\frac{t}{t_{\rm vi}})^{1/2},
\end{equation}
where, $\gamma^{2}=r_{\rm i}/a$, $a$ is the binary separation, and
$t_{\rm vi}$ the viscous timescale at the inner edge of the CB
disk,
\begin{equation}
t_{\rm vi}=\frac{4r_{\rm i}^{2}}{3\nu_{\rm i}},
\end{equation}
where $\nu_{\rm i}$ is the viscosity at the inner edge of the CB
disk. We estimate $\nu_{\rm i}$ using the standard $\alpha$
prescription \citep{Shak73},
\begin{equation}
\nu_{\rm i}=\alpha_{\rm SS} c_{\rm s} H_{\rm i},
\end{equation}
where $c_{\rm s}$ and $H_{\rm i}$ are the sound speed and the
height of the disk at the inner edge of disk, respectively. In the
following calculations we set the viscosity parameter,
$\alpha_{\rm SS}=0.001$, and assume that the disk was
hydrostatically supported and geometrically thin with $H_{\rm
i}/r_{\rm i}\sim 0.03$ \citep{bell04}. The sound speed can be
obtained from the equation of vertical hydrostatic equilibrium,
\begin{equation}
c_{\rm s}\simeq\Omega_{\rm K} H_{i},
\end{equation}
where $\Omega_{\rm K}=(GM_{\rm T}/r_{\rm i}^{3})^{1/2}$ is the
Keplerian angular velocity at $r_{\rm i}$.

\noindent{\em 3. the gravitational and baryonic mass of the
neutron star}

For neutron stars, one should consider the discrepancy between its
baryonic mass, $M_{\rm b}$, and the gravitational mass, $M_{\rm
g}$. By differentiating the binding energy equation
(\cite{latt89}),
\begin{equation}
M_{\rm b}=M_{\rm g}+0.084M_{\odot}(\frac{M_{\rm
g}}{M_{\odot}})^{2},
 \end{equation}
with respect to $t$, the relation between the growing rate of the
gravitational and of the baryonic masses is
\begin{equation}
\dot{M}_{\rm X}=\dot{M}_{\rm g}=(1-\beta)\dot{M}_{\rm b},
\end{equation}
where $\beta=1-(1+0.168M_{\rm X}/M_{\odot})^{-1}$. We assume that
the fraction $\beta$ of the accreted baryonic mass disappears
along with the specific angular momentum of the neutron star.

\subsection{Computing $\xi_{R}$}

Considering various kinds of mass loss discussed in the above
subsection, we can write the rate of change of the orbital angular
momentum of the binary to be
\begin{equation}
\frac{\dot{J}}{J}=[(1-\delta)\frac{M}{M_{\rm T}M_{\rm X}}+\eta
\frac{\gamma}{\mu}]\alpha \dot{M}+\frac{M}{M_{\rm T}M_{\rm
X}}(1-\alpha)\beta\dot{M},
\end{equation}
where $\eta=\delta(t/t_{\rm vi})^{1/2}$ is the efficiency factor
of transfer mass removing angular momentum from the system through
the CB disk; $\mu=MM_{\rm X}/M_{\rm T}$ is the reduced mass of the
binary. The first, second, and third terms on the right-hand side
of equation (17) represent the change in the orbital angular
momentum due to mass loss through ejected outflows, the CB disk
and the baryonic mass loss, respectively.

For $M \leq 0.8 M_{\rm X}$, the Roche lobe radius of the donor can
be approximately denoted by (\cite{pacz71})
\begin{equation}
R_{\rm L}=0.462a\left(\frac{q}{1+q}\right)^{1/3},
\end{equation}
where $q=M/M_{\rm X}$. We then obtain $\xi_{\rm R}$ combining
equations (7), (17), and (18):
\begin{eqnarray}
\xi_{\rm R} & = &-\frac{5}{3}+2(1-\alpha)(1-\beta)
 q+\frac{2}{3}(\alpha+\beta-\alpha\beta)\frac{q}{1+q}+ \nonumber\\
& & 2[\alpha(1-\delta)+\beta(1-\alpha)]\frac{q^{2}}{1+q}
 +2\gamma\eta\alpha(1+q).
\end{eqnarray}
When $\alpha=\beta=0$, equation (19) recovers to the standard
conservative mass-transfer case,
\begin{equation}
\xi_{\rm R}=-\frac{5}{3}+2q.
 \end{equation}
When $\beta=\delta=0$, equation (19) becomes
\begin{equation}
\xi_{\rm R}=-\frac{5}{3}+2(1-\alpha)q+\left(\frac{2}{3}+2q\right)
\frac{\alpha q}{1+q}
 \end{equation}
for nonconservative mass transfer with only outflows inclued
(\cite{Lixd98}).

\section{Numerical Results}
We adopted the semi-analytical method by Webbink et al. (1983) to
calculate the evolutionary sequences of LMXBs. We set the initial
mass of the donor $M=1.0M_{\odot}$ (with solar chemical
composition X=0.7, Y=0.28) and the core mass in the range of
$0.15M_{\odot}\leq M_{\rm c}\leq 0.35M_{\odot}$, the initial mass
of the accreting neutron star being $M_{\rm X}=1.4M_{\odot}$.

For stable mass transfer, we let $\dot{M}_{\rm
X}=-(1-\alpha)(1-\beta)\dot{M}$ when $-\dot{M}>\dot{M}_{\rm E}$.
When the accretion disk becomes unstable, we calculate $\Delta
M_{\rm X}$ according to equation (8). For the radius $r_{\rm i}$
of the inner edge and the mass feeding rata $\delta$ of CB disk,
we took $r_{\rm i}/a=\gamma^{2}=1.7$ (\cite{arty94}) and
$\delta=0$, 0.01 and 0.02. We stopped the calculations when the
mass of the hydrogen envelope of the giant was reduced to be
$\sim2\%$ of the core mass.

Figure 1 shows the calculated $\xi_{\rm R}$ against the mass
ratio, $q=M/M_{\rm X}$, for the initial $M_{\rm c}=0.25 M_{\odot}$
and $\delta =0$, 0.01, 0.02. One can see that the larger is
$\delta$, the smaller is $-\xi_{\rm R}$, and hence the higher the
mass transfer rates. This can also be seen in Figure 2, which
compares the detailed evolutions of the the giant mass, the
orbital period, the neutron star mass, and the mass-transfer rate
for different values of $\delta$. Obviously, the existence of the
CB disk can accelerate the evolutionary processes quite
efficiently. If $\delta$ is further increased, the orbital period
will decrease, finally leading to runaway mass transfer. The
variation of $\dot{M}$ is also much larger if the CB disk effect
is included. Because of unstable mass transfer induced by the
accretion-disk instability, the efficiency of neutron-star
accretion is quite low. The resulting averaged accretion rates of
the neutron star, $\dot{M}_{\rm X}=\Delta M_{\rm X}/t_{r}$, are
shown in Figure 3.

\section{Application to PSR J1713+0747}

We have performed many calculations for LMXBs evolution with
different initial parameters in order to fit the observed data
($P_{\rm orb}$, $M_{\rm X}$, and $M$) of PSR J1713+0747. We found
reasonable results when the initial core mass was
$0.258M_{\odot}$, $0.246M_{\odot}$, $0.205M_{\odot}$ for
$\delta=0.02$, 0.01, 0, respectively, while conservative evolution
failed to produce results compatible with all of the measured data
simultaneously. Table 1 summarizes the results of our
calculations. It seems that both models with and without a CB disk
can well fit the observed data for this source, if unstable mass
transfer is taken into account. However, there exist considerable
differences in the initial binary parameters and the evolutionary
timescales, suggesting that the CB disks may have important
implications for the LMXB evolution (see below).

Previous investigations suggested a simple relation between the
orbital period, $P_{\rm orb}$, and the mass, $M$, of the white
dwarf for low-mass binary radio pulsars. Assuming the mass-feeding
rate to be $\delta=0.002$ (a larger $\delta$ possibly cause
infinite mass transfer rate when initial core mass $M_{\rm
c}=0.15M_{\odot}$, see Figure 1) of the CB disk, we calculated 21
LMXBs evolution sequences, where the initial core mass of giant
ranged from 0.15 to $ 0.35M_{\odot}$. The relation between $P_{\rm
orb}$ and $M$ when evolution ceases for $0.15M_{\odot}\leq M_{\rm
c, i}\leq 0.35M_{\odot}$ was plotted with the solid curve in
Figure 4; the dot and dashed curves correspond to the relations
obtained by \citet{rapp95} and \citet{taur99}. It can be seen that
there is a power-law relation between $P_{\rm orb}$ and $M$ with
$P_{\rm orb} ({\rm d})\propto (M/M_{\odot})^{6.55}$.

\section{Discussion and Summary}

The mass measured for PSR J1713+0747 indicates that the neutron
star had accreted $< 0.1 M_{\odot}$ mass during its previous LMXB
evolution, if formed with an initial mass of 1.4 $M_{\odot}$. We
suggest that unstable mass transfer in the accretion disk may lead
to extensive mass loss in the evolution of relatively wide LMXBs.
The mass accreted by the neutron star is considerably smaller than
in the stable accretion case (see also \cite{Lixd98}).

The evolutionary path of the binary also depends on the form of
mass and angular momentum loss when mass transfer is
nonconservative. Traditionally the excess material is assumed to
leave the binary with a specific orbital angular momentum of
either the neutron star or the binary. However, based on arguments
proposed by Spruit and Taam (2001) and Taam and Spruit (2001), we
considered the case with a CB disk originating from part of the
mass outflow. Actually, radio observations of super-Eddington
accretion systems SS 433 \citep{blun01} and Cygnux X-3
\citep{mill04} have shown evidence of equatorial, disk-like
outflows.

Our calculations show that the CB disk can remove the orbital
angular momentum of the binary at a rate significantly higher than
simple mass outflows. This may have important implications on the
LMXBs evolution and its relation with low-mass binary pulsars.
Recent binary population synthesis studies \citep{pods02,pfah03}
fail to produce enough luminous LMXBs, as observed. This problem
seems to be related to the discrepancy between the birthrates of
LMXBs and low-mass binary pulsars (e.g. \cite{kul88}), the
solution of which may lie in an unknown mechanism for
angular-momentum loss to speed up the mass transfer. Moreover, it
has been pointed out by \citet{sill00} that the traditional
magnetic braking, which has served as the basis for the evolution
of LMXBs and cataclysmic variables (CVs), is significantly
reduced. The CB disk may provide an efficient mechanism to drain
the orbital angular momentum, accelerating the binary evolution
and enhancing the mass-transfer rates. \citet{dubu02} found that
the spectral energy distributions expected of CB disks in CVs can
dominate the emission from the donor star and the accretion disk
of the white dwarf at wavelengths $\gsim 3 \quad\mu$m. At longer
wavelengths the relative contribution from the CB disk to the
total emission from the system increases. Since LMXBs and CVs have
similar binary parameters, searching for (mid- and far-) infrared
emission from the CB disks will be of great importance for the
evolution of both LMXBs and CVs.
\\

We are grateful for an anonymous referee for helpful comments.
This work was supported by the Natural Science Foundation of China
(NSFC) under grant number 10573010.

\clearpage
\begin{table}
\caption{ Calculated results for different evolutionary models. *
}
\begin{center}
\begin{tabular}{llllllll}
\hline\noalign{\smallskip}
\hline\noalign{\smallskip}
Model&$M_{\rm c}(t_{0})$&$t_{\rm f}-t_{0}$&$\rm log|\dot{M}(t_{\rm f})|$&$R(t_{\rm f})$&$P_{\rm orb}(t_{\rm f})$&$M_{\rm c}(t_{\rm f})$&$M_{\rm X}(t_{\rm f})$ \\
$(\delta,\alpha,\beta)$&$(M_{\odot})$ & (Myr) &$(M_{\odot}$yr$^{-1})$&$(R_{\odot})$ & (days)& $(M_{\odot})$&$(M_{\odot})$\\
\hline\noalign{\smallskip}
$\delta=0.02 $&0.258&23.9& $-7.95$&21.38& 68.36&0.284 &1.402\\
$\delta=0.01$&0.246&47.1&  $-8.15$&21.20  & 67.71& 0.283& 1.403\\
$\delta=0 $&0.205&267.5& $-8.406$&21.67& 69.78&0.284 &1.408\\
$\alpha=\beta=0$&0.215&179.00&$-8.400$&20.99&66.70&0.283&2.117\\
\noalign{\smallskip}\hline
\end{tabular}\\
\noindent{* $t_0$ and $t_f$ denote the ages at the beginning and
end of the mass transfer, respectively.}
\end{center}
\end{table}

\begin{figure}
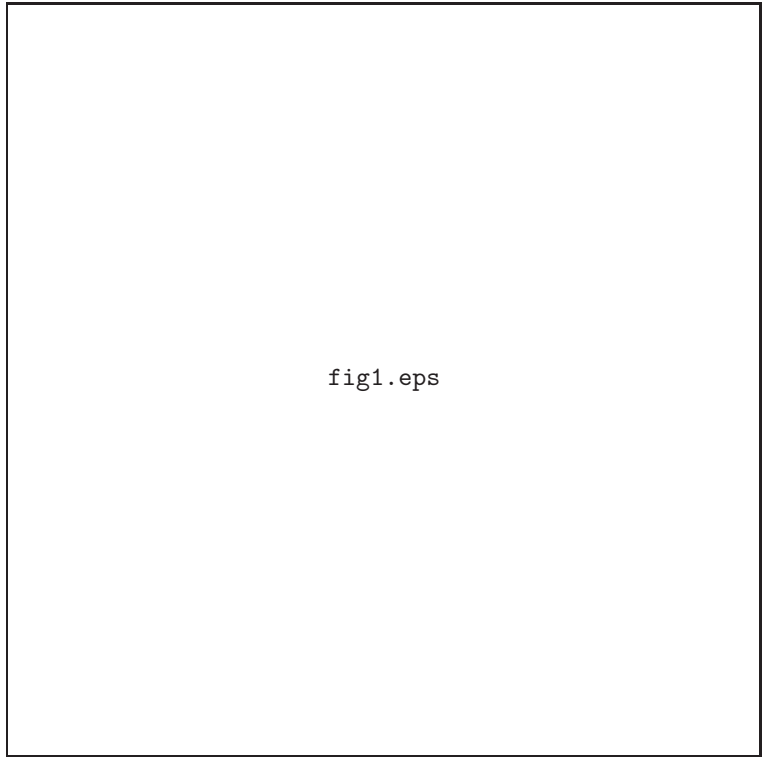

\begin{center}
\FigureFile(100mm,100mm){fig1.eps}
\end{center}
\caption{Plot of $\xi_{\rm R}$ vs. $q$ for LMXBs with an initial
core mass of $M_{\rm c} = 0.25 M_{\odot}$. The dot, dashed, and
solid curves correspond to $\delta = 0$, 0.01, and 0.02,
respectively.}\label{fig1}
\end{figure}

\clearpage
\begin{figure}
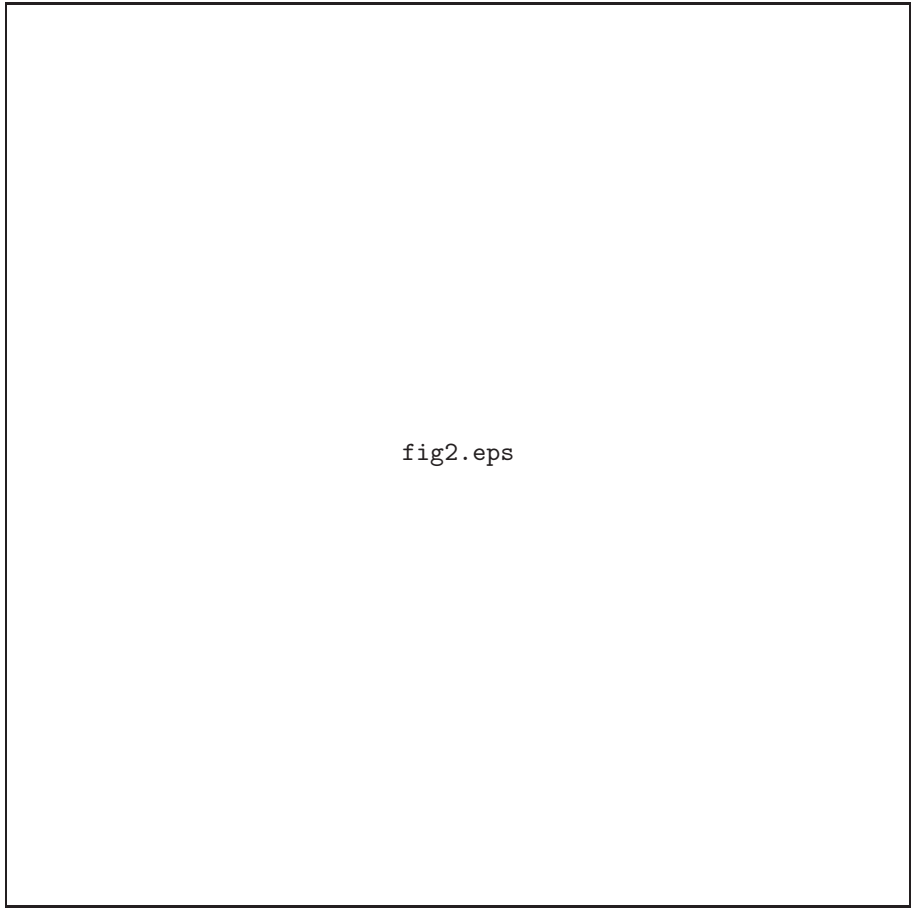

\begin{center}
\FigureFile(120mm,120mm){fig2.eps}
\end{center}
\caption{Evolutions of an LMXB with an initial donor mass of
$M=1.0M_{\odot}$ and a core mass of $M_{\rm c}=0.25M_{\odot}$. The
dot, dashed, and solid curves correspond to $\delta = 0$, 0.01,
and 0.02, respectively.}\label{fig2}
\end{figure}

\begin{figure}
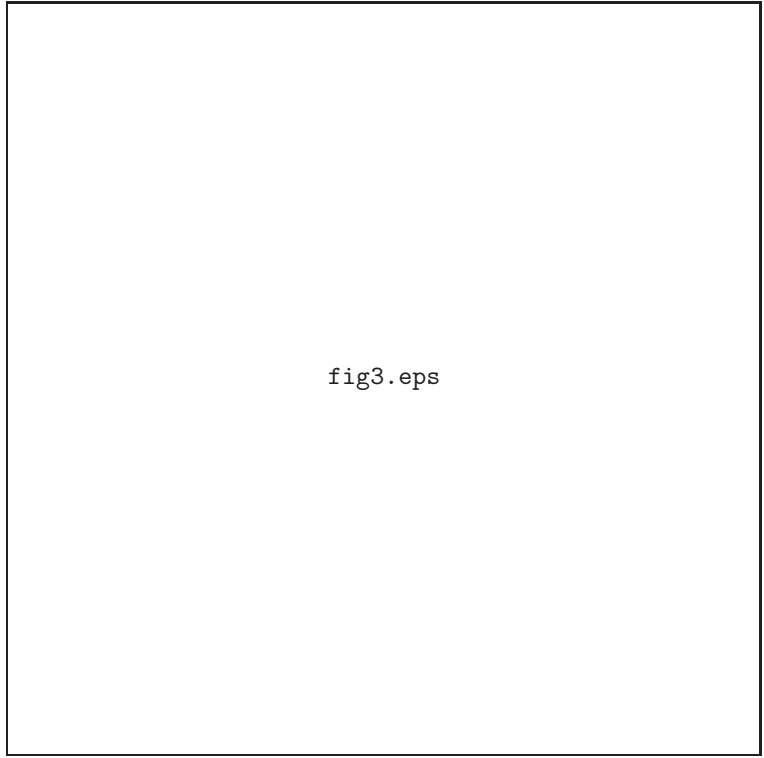

\begin{center}
\FigureFile(100mm,100mm){fig3.eps}
\end{center}
\caption{Evolution of the accretion rate of the neutron star (mean
accretion rate in a recurrence time $t_{\rm r}$).  The dot,
dashed, and solid curves correspond to $\delta = 0$, 0.01, and
0.02, respectively.}\label{fig3}
\end{figure}

\clearpage

\begin{figure}
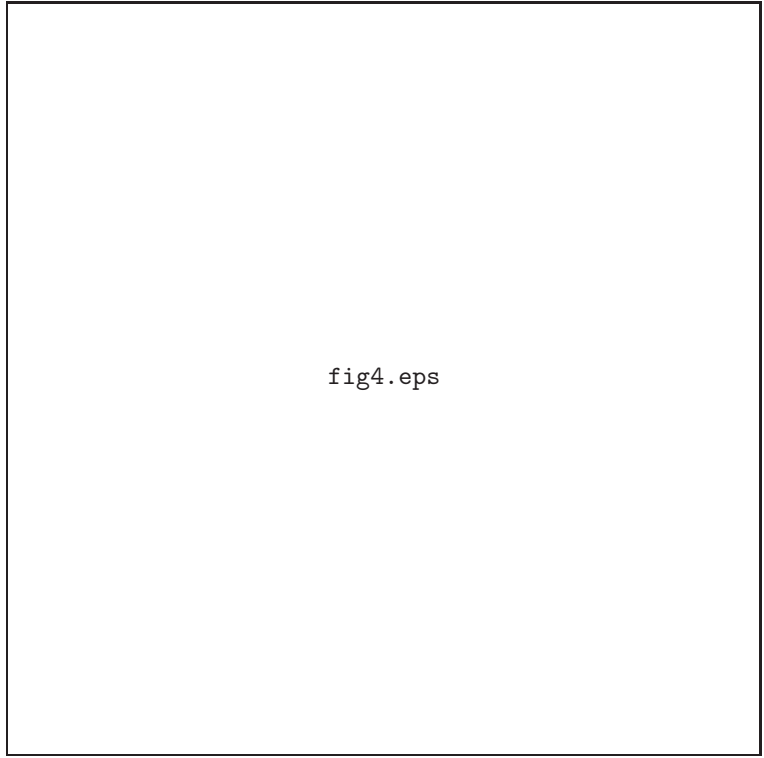

\begin{center}
\FigureFile(100mm,100mm){fig4.eps}
\end{center}
\caption{Predicted relation between $P_{\rm orb}$ and $M$ for
low-mass binary radio pulsars. The dashed, dot, and solid curves
correspond to the relations obtained by \citet{taur99},
\citet{rapp95}, and this work, respectively.}\label{fig4}
\end{figure}

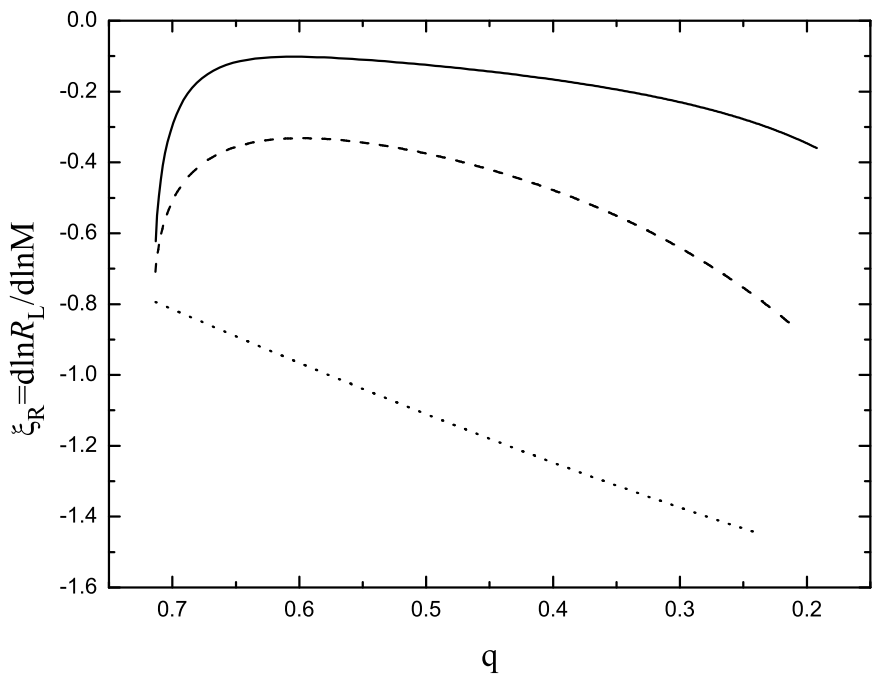
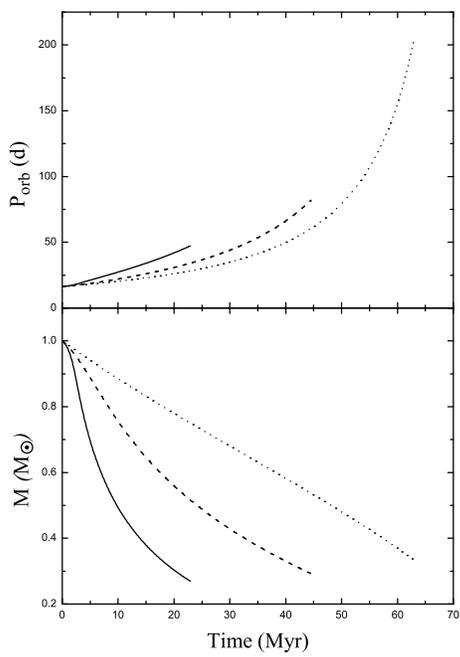
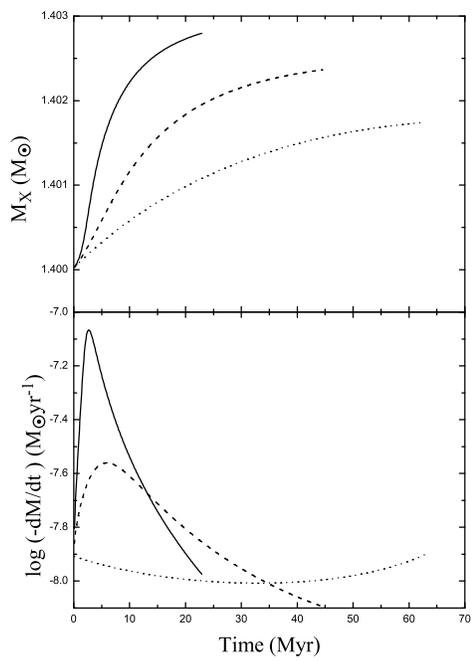
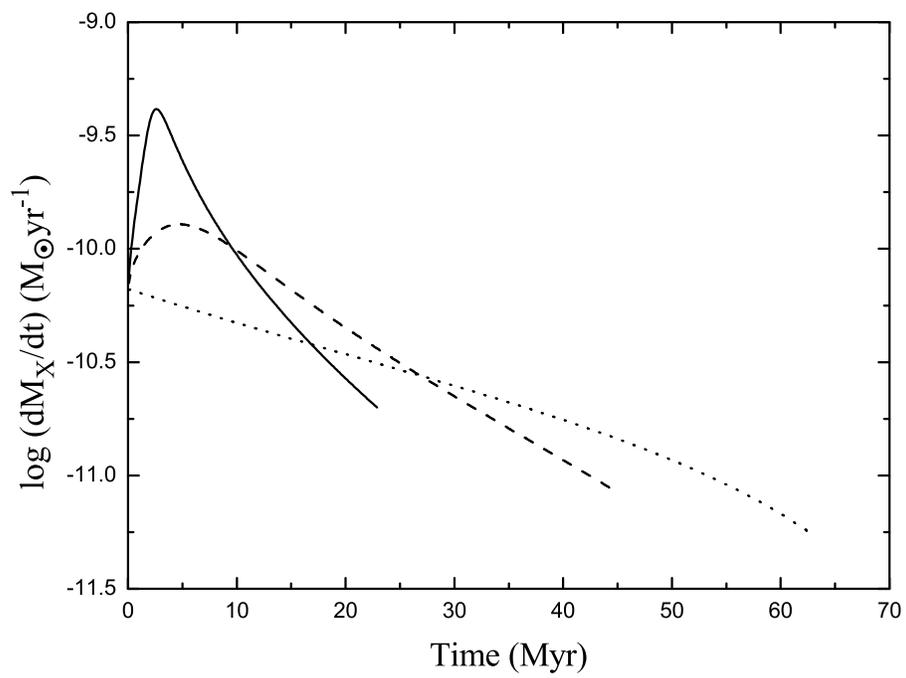
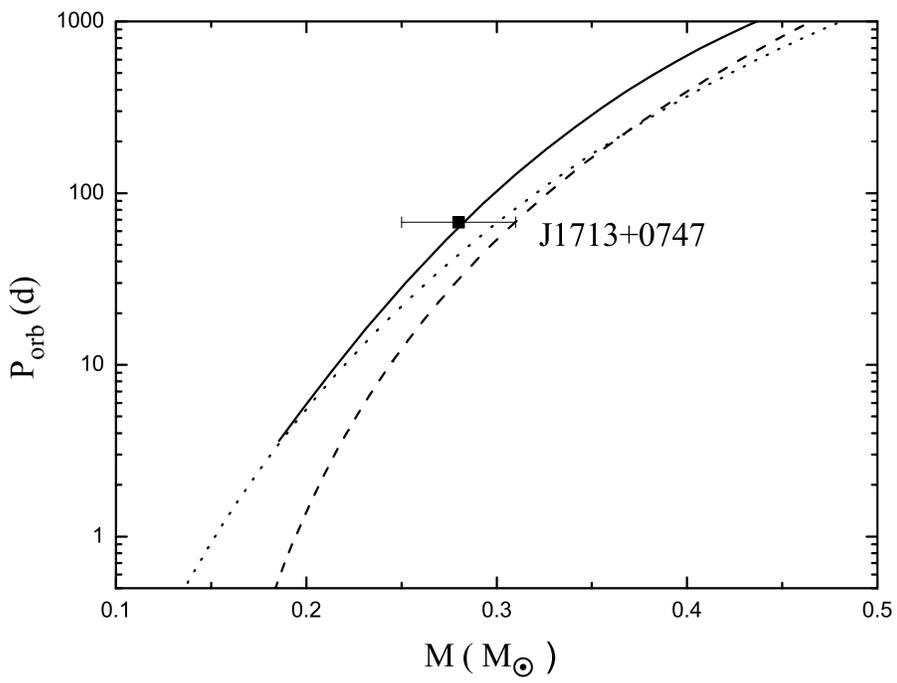
\end{document}